# Assessing the Impact of Flip Angle and on Image Quality and reliability of Ernst Angle optimization across Varied Conditions in Magnetic Resonance Imaging

Zhiyi Ding

# Table of Contents

## Abstract

This study investigates the significance of flip angle, an imaging parameter, in enhancing Magnetic Resonance image quality under various imaging conditions. It specifically explores the extent to which the Ernst angle, an optimal flip angle, optimizes image quality under different imaging parameters. The investigation begins with a theoretical derivation of the Ernst angle, assuming steady state imaging conditions. Then multiple studies that examine the effect of flip angle on signal-to-noise ratio (SNR), a key indicator of image quality, in different areas of the human body (blood, liver, and brain), are analysed. The study compares the results of these studies and compares their respective optimal flip angles with the Ernst angle. The findings reveal that flip angle plays a crucial role in enhancing SNR and image quality. However, the Ernst angle only optimizes SNR under steady state conditions and when using a spoiled gradient echo (GRE) sequence. Therefore, further investigations are necessary to determine the optimal flip angle under different imaging conditions to optimize SNR and enhance overall image quality.

## Introduction

Magnetic Resonance Imaging (MRI) is a non-invasive diagnostic imaging technique that investigates a patient's body anatomy, lesions, and organ function based on the magnetic properties in atoms. In this study, the significance of flip angle on image quality and the reliability of Ernst angle optimization is explored. It is imperative to have high-quality MRI images for accurate diagnosis by medical professionals and exploring the efficacy of Ernst angle optimization will aid in determining a more efficient approach for optimizing flip angles during clinical practice.

## Background Theory

### MRI structure

The MRI system mainly consists of a magnet, Radiofrequency (RF) coils, Gradient coils, and a computer system.

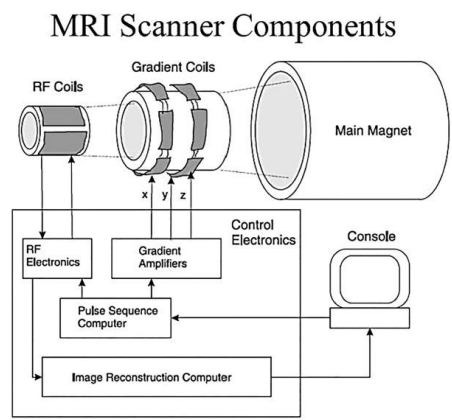

Figure 1, MRI scanner components, by Gruber, 2018

In figure 1, the patient aperture is surrounded by the RF coils that emit radio waves and receive signals to and from the samples. Then, it is surrounded by gradient coils, which generate secondary magnetic fields that allow spatial localisation of the image. It is further surrounded by the main magnet, commonly a superconducting magnet, that generates a strong uniform magnetic field, $B_0$. Collected analogue data is converted into digital data by connecting electronics to the computer system.

### Nuclear Magnetic Resonance (NMR)

When placed in an external magnetic field, nuclei absorb electromagnetic radiation at their resonant frequencies.

### Proton spin and precession

Protons, or in general atomic nuclei, acts as gyroscopes that spin about their axis, generating a spinning magnetic field due to electromagnetic induction known as magnetic moment. For instance, a gyroscope on Earth precesses, or 'wobbles', due to the Earth's gravitational field. This precession is a result of the interaction between the angular momentums produced by the spinning gyroscope and the Earth. The same logic can be applied to protons. In the presence of an external magnetic field $B_0$, protons precess perpendicularly to the direction of $B_0$. This motion is demonstrated in figure 2,

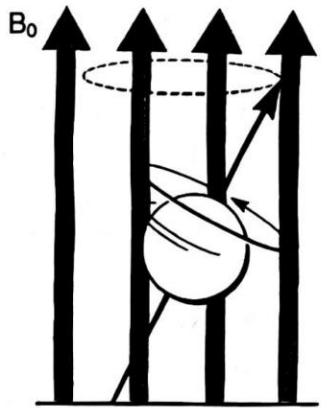

Figure 2, proton precession under $B_0$, by Bushong, 2015

### Larmor Frequency

Larmor Frequency is the rate at which nuclear species precess under $B_0$. It is essential to target the sample with Radiofrequency pulses (RF pulses) at Larmor Frequency as energy transfer is most efficient when at resonance, allowing protons to be excited.

### Signal Detection

During imaging, firstly, the sample (patient) will be placed into the aperture, under the strong magnetic field, $B_0$, and align with the direction of the external field in equilibrium (Pope, 1999, pp.76-77). The net magnetisation of all individual magnetic moments of the system (Bushong, 2015, pp.43), $M$, is conventionally considered to lie in the Z direction in a 3D frame, as demonstrated in figure 3:

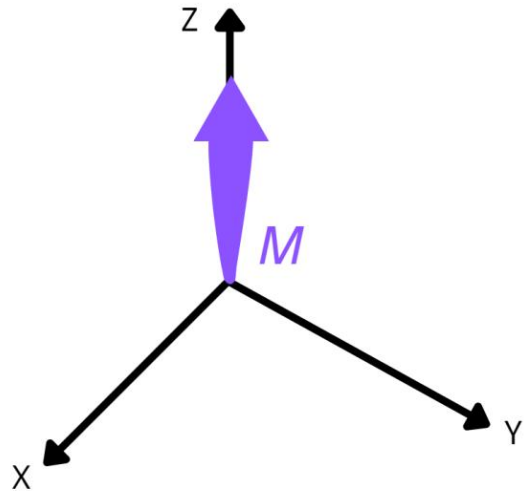

Figure 3, direction of *M* in a 3D frame, by self

Then, Radiofrequency (RF) Pulse, a small oscillating magnetic field at the specific Larmour frequency, is applied perpendicular to B$_0$ toward the specie (Navigating Radiology, 2022). This tips the *M* towards the XY plane and allows the nuclei to precess at this frequency. In practice, various RF pulses that tips *M* to various angles away from the Z-axis. For example, a 90° RF pulse tips *M* parallel to the XY plane, as demonstrated in figure 4:

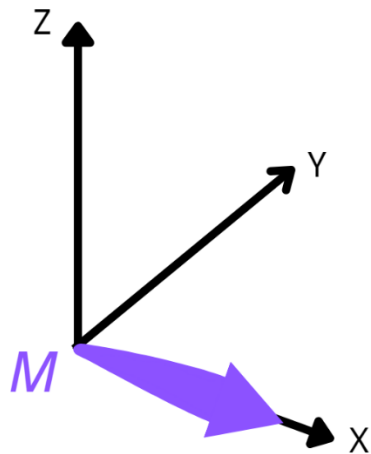

Figure 4, *M* tipped 90° , by self

This tipping generated a transverse magnetic field only in this 90° pulse incidence. However, in general, the net magnetization has two components: longitudinal (*M$_z$*) and transverse magnetisation (*M$_{xy}$*). The oscillation of transverse magnetic field is the sole source of MRI signal due to the construction of the MRI machine, as demonstrated in figure 5:

Acronyms:

*M*: net magnetization, M$_{xy}$: strength and direction of transverse magnetic field, M$_z$: strength and direction of longitudinal magnetic field, B$_0$: external background magnetic field strength, TR: repetition time, SNR: signal-to-noise ratio

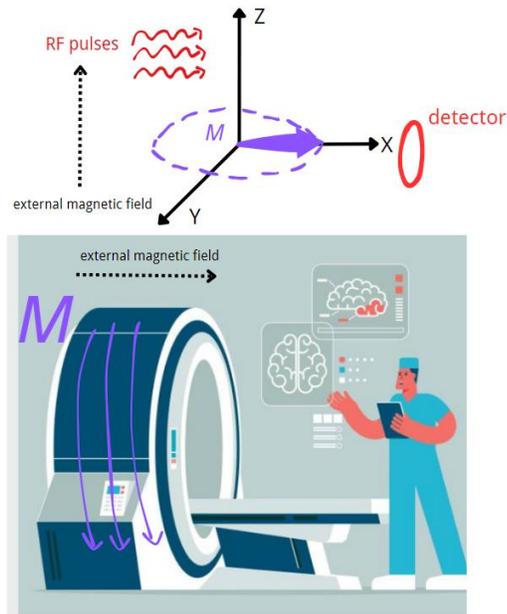

Figure 5, bottom image by istock, 2022

After excitation, nuclei seek to realign with the direction of M₀, releasing energy and and returning to equilibrium. The longitudinal magnetic field recovers exponentially during T1 relaxation time, a quantity that measures the time for this recovery and the transverse magnetic field returns to equilibrium during T2 relaxation time, which is generally shorter than T1 (Pope, 1999, pp.73). In the RF coil, a current is detected from the oscillating and decaying transverse magnetic field through electromagnetic induction, while the longitudinal magnetization, parallel to the coil, does not contribute to the signal. Hence, the detectable signal originates from the transverse magnetic field in the XY plane.

### *Flip Angle*

Flip angle a imaging parameter that rotates angle of net magnetisation away from the Z-axis after nuclei absorbed energy from RF pulse, countering B₀ (Bushong, 2015).

It is in 1984 that the desire to develop faster imaging had grown significantly. A new group of pulse sequence, which lacks the 180° RF pulse that is used in Spin Echo (SE) imaging, is introduced. The removing of the 180° RF pulse allowed repetition time (TR), the time between each RF pulse sequence, to shorten.

The advancement in fast imaging was introduced in 1984 with a pulse sequence that eliminated the 180° RF pulse used in Spin Echo (SE) imaging, allowing for shorter repetition times (TR) between RF pulse sequences.

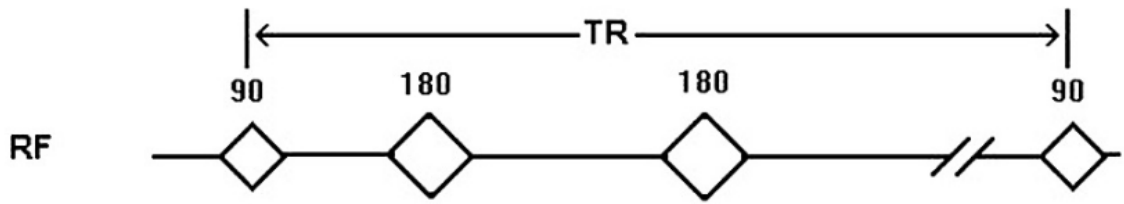

Figure 6, repetition time in a pulse sequence, by Ballinger

The reduction in TR further decreases the range of $M_z$, which reduces the SNR. A mitigate this issue, the flip angle needs to be adjusted. It is is optimised with TR and T1 relaxation time (time for longitudinal magnetisation to return to equilibrium) (Bushong, 2015).

*A Good Image*

The highest quality achievable from any MRI is crucial in medicine, which involves optimizing image resolution to ensure adequate image visibility and enabling clear differentiation among tissues. In addition, the images must be precise and representative. Achieving these standards is essential for the images to be considered suitable for clinical diagnosis.

This investigation will be focusing on one indicator of image quality: the signal-noise-ratio (SNR).

During imaging, the MRI signal is converted into a digital value that represents pixel intensity, known as the signal, while unwanted background and electronic noise recorded during conversion are detected as noise. SNR qualifies image quality by comparing signal intensity and noise level. This can be calculated as follows:

$$SNR = \frac{Signal\ intensity}{Noise\ level}$$

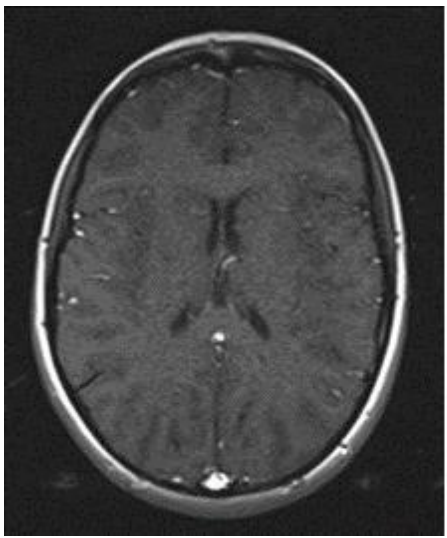

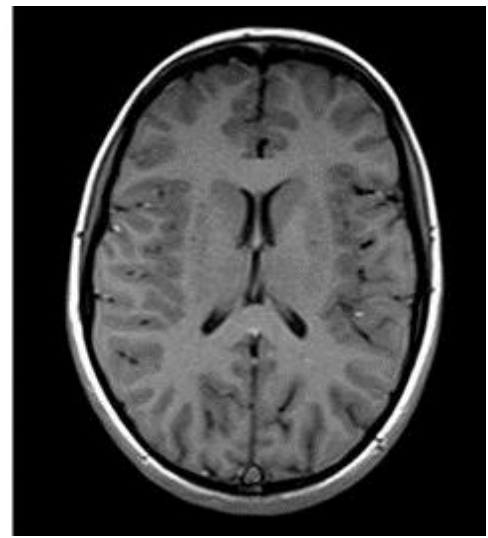

**Low SNR**                    **High SNR**

Figure 7, A comparison between images with high and low SNR, by mrimaster, 2023

Acronyms:

*M*: net magnetization, $M_{xy}$: strength and direction of transverse magnetic field, $M_z$: strength and direction of longitudinal magnetic field, $B_0$: external background magnetic field strength, TR: repetition time, SNR: signal-to-noise ratio

Higher SNR improves image quality and brightness, while a larger signal intensity contributes to a greater SNR. Similarly, lower noise levels increase image visibility. SNR is a vital metric in imaging, providing insights into image quality for lesion diagnosis, contrast enhancement measurement, and image evaluation by doctors and radiologists.

To increase SNR, the general approach involves increasing signal strength or reducing background noise. In this investigation, the signal strength will be maximised by modifying the flip angle while maintaining constant noise levels, thereby enhancing SNR.

## Theoretical Analysis:

The derivation of optimising the flip angle under steady state builds upon C. Olman's work, 2020, and applies a mathematical framework to investigate the behavior of magnetization in MRI. The Bloch equation, which describes magnetization under a magnetic field, is analyzed to determine the dynamics of $M_z$ and $M_{xy}$. An inversion recovery experiment is conducted to examine magnetization changes before and after an RF pulse. Equations are derived to capture the decay of $M_z$, $M_{xy}$ at a given flip angle, and $M$ in steady state. By optimizing the flip angle, signal intensity can be maximized. The study emphasizes the mathematical analysis of magnetization dynamics in MRI.

In summary, the optimum flip angle, the Ernst angle, for MRI under a steady state condition can be calculated as:

$$\alpha = cos^{-1}(e^{-TR/T_1})$$

Where $\alpha$ stands for flip angle, TR stands for repetition time, and T1 stands for T1 relaxation time.

See full derivation in appendix.

## Practical Analysis

### Study 1:

In 2011, Ennis conducted a study on the femoral arteries in the thigh using phase-contrast MRI (PC-MRI) with gradient echo sequencing, to explore the correlation between flip angle and SNR. The study's importance stems from the prevalent utilization of suboptimal flip angles in PC-MRI, despite the previous research highlighting the enhanced SNR possibility using optimized flip angles (Gao, 1988).

The study involved five healthy volunteers placed in a 1.5 Tesla magnetic field, with measurements taken five times with each flip angle values 5°, 10°,15°,20°,30°,40°,60°,75° and 90°. Then, a graph of SNR against Flip angle is plotted:

Acronyms:

$M$: net magnetization, $M_{xy}$: strength and direction of transverse magnetic field, $M_z$: strength and direction of longitudinal magnetic field, $B_0$: external background magnetic field strength, TR: repetition time, SNR: signal-to-noise ratio

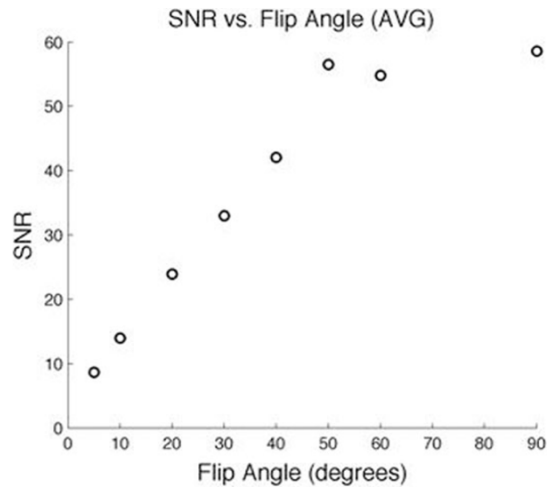

Figure 8, relationship between SNR and Flip Angle, by Ennis, 2011

Figure 8 demonstrates a noteworthy positive correlation between the Signal-to-Noise Ratio (SNR) and the Flip Angle. The reason for this phenomenon is that a greater amount of *M* is oriented towards the XY plane, resulting in an increased in $M_{xy}$. As a result, the signal intensity is amplified correspondingly. At the outset, it can be observed that there exists a positive correlation between the Flip Angle and SNR within the range of 10° to 40°. Nevertheless, once the angle exceeds 40°, there is a nonlinear relationship, and the SNR at 50° manifests an anomaly, which is attributable to the small sample size. However, it is noteworthy that beyond a 40° angle, the SNR increase decelerates in tandem with the decline in the signal-to-noise ratio (SNR). This phenomenon can be attributed to the non-linear positive correlation between flip angle and signal intensity, as clearly depicted in Figure 9:

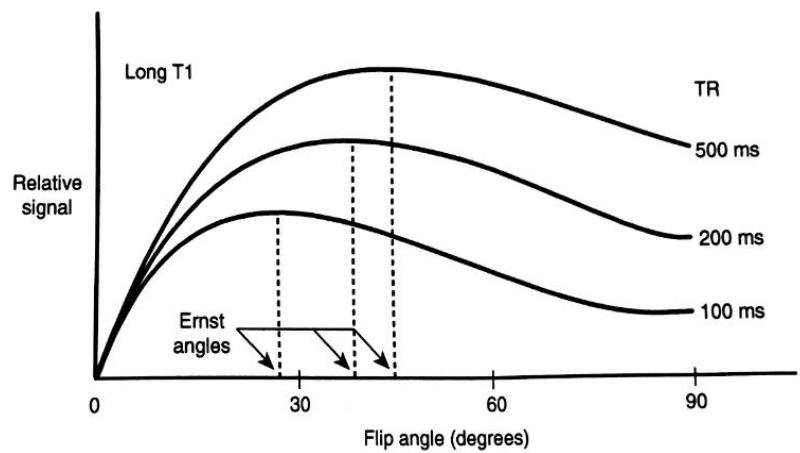

Figure 9, relationship between flip angle and relative signal at various TR, by Bushong, 2015, pp281

As such, a non-linear correlation can be observed between the flip angle and SNR, which aligns with the pattern seen in the relationship between the flip angle and signal intensity. It is important to note that the SNR exhibits a range of values from approximately 10 to 60, demonstrating a 500% increase, which is substantial.

| Tissue | Approximate T1 Relaxation Times (in ms) at Three $B_0$ Field Strengths for Various Tissues | | |
|---|---|---|---|
| | 0.35 Tesla | 1.5 Tesla | 3.0 Tesla |
| Adipose | 200 | 260 | 290 |
| Liver | 285 | 495 | 640 |
| Renal cortex | — | 1000 | 1150 |
| Renal medulla | — | 1300 | 1550 |
| Spleen | 485 | 780 | 985 |
| White matter | 500 | 650 | 800 |
| Gray matter | 800 | 1200 | 1500 |
| Muscle | 500 | 1000 | 1400 |
| Heart muscle | 525 | 870 | 1100 |
| Oxygenated blood* | 900 | 1350 | 1650 |
| Deoxygenated blood* | 900 | 1350 | 1575 |
| Cerebrospinal fluid | 3300 | 3350 | 3400 |
| Water | 3500 | 3500 | 3500 |

*At typical human hematocrit of 0.42.

Figure 10. A table with T1 values of various tissues at different magnetic field strengths, by Bushong, 2015, pp.68.

Acknowledging the fact that oxygenated blood demonstrates a relatively high T1 value of 1350ms in figure 10 in a magnetic field of 1.5T, it is unexpected to observe that a large flip angle is at optimum in practice despite the expectation of a small optimum flip angle set by the Ernst angle equation.

*Study 2:*

In a comprehensive study conducted by Lee in 2015, the hepatic parenchyma (the liver tissue) of 63 patients diagnosed with Hepatocellular Carcinoma (HCCs), a type of liver cancer, were imaged using Gradient Echo (GRE) pulse sequences under a 1.5 Tesla background magnetic field. For each patient, a series of images was acquired using successive flip angles of 10°, 20°, and 30°. Subsequently, signal intensity calculations were performed by selecting a consistent, large region of interest (ROI) at an identical location with all images. This approach ensured standardised analysis across all patients. Additionally, noise intensity was estimated by measuring the standard deviations of noise signals measured outside the body, with this measurement repeated three times and an average taken. Finally, a box-and-whisker plot was derived from the combined dataset, incorporating data collected from all the participating patients:

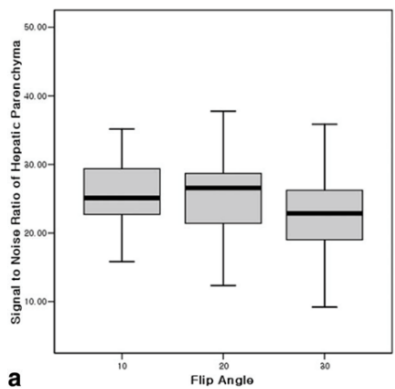

Figure 11. A box-and-whisker plot of SNR of Hepatic Parenchyma against Flip angle, by Lee, 2015

Acronyms:

*M*: net magnetization, $M_{xy}$: strength and direction of transverse magnetic field, $M_z$: strength and direction of longitudinal magnetic field, $B_0$: external background magnetic field strength, TR: repetition time, SNR: signal-to-noise ratio

In Figure 11, the SNR reveals an intriguing trend. It initially increases from 10° to 20° as signal intensity rises due to greater magnetization tipping into the transverse plane, resulting in increased $M_{xy}$ and hence MRI signal. However, between 20° and 30°, the SNR decreases even below the value at 10°. Although there is a larger $M_{xy}$, $M_z$ for the subsequent excitation is diminished (Bushong, 2015, pp.282). This results in a significant reduction in the magnitude of the following transverse magnetic field, leading to decreased signal intensity.

The plotted data exhibits an asymmetrical distribution across all variables except the flip angle at 30 degrees. The observed disparity may be due to the diverse demographic characteristics of the volunteers, leading to differences in their hepatic tissue. As the flip angles increases, the range of data expands. Notably, the median SNR reaches its peak at 20° with a value of 28.33, followed by 25.33 at 10° and 23.67 at 30°. Thus, it can be inferred that the maximum increase in SNR is within 30° to 20°, yielding a 20% elevation increase. In conclusion, the study recommended the utilization of a 20° angle to optimize SNR.

Furthermore, the significant increase in flip angle changes leads to considerable uncertainties between each angle. Hence, it can be inferred that the existing evidence is inadequate to conclusively establish the optimal angle as 20° but rather between 20° and 30°, contrary to what was suggested by the study.

Importantly, liver tissue at a magnetic field intensity of 1.5 Tesla exhibits an exceptionally low T1 value of 495 milliseconds, second only to adipose tissue. Considering this information and the Ernst Angle formula, one would expect an increased flip angle to be optimal. However, the actual outcome is unexpected.

*Study 3:*

Tanaka's (2009) research utilised MRI imaging to measure the SNR of the bilateral internal carotid artery and basilar artery situated within the neck. The SNR was evaluated through the manipulation of flip angles, within a range spanning from 4° to 90°. A total of 40 volunteers were included for flip angles of 4°, 15°, 60°, and 90°, while 54 volunteers were involved for flip angles of 8°, 15°, and 30°. Sequential image acquisition was performed in increasing order of flip angles within a 1.5 Tesla magnetic field. SNR and signal intensity were measured using consistent regions of interest (ROI) across the samples. The resulting SNR values were recorded in a table for analysis.

To visually represent the dataset used in this investigation, a line graph is plotted:

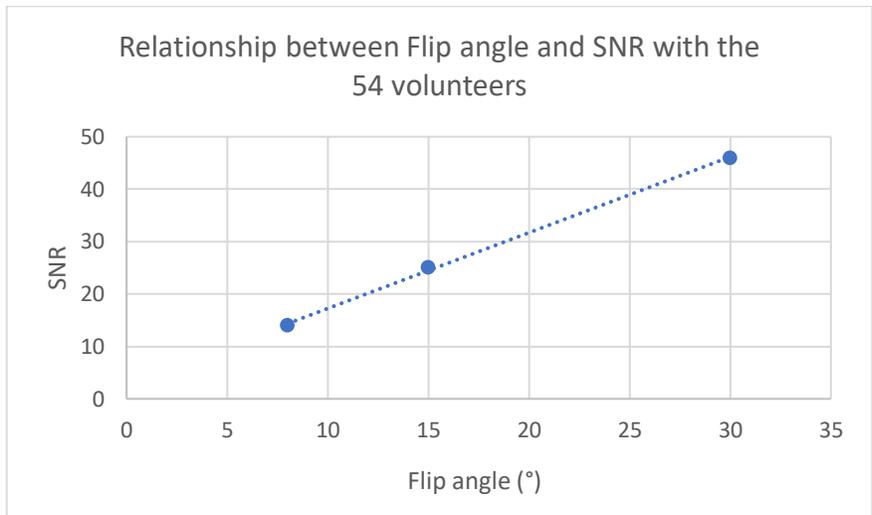

Figure 12. Dataset by Tanaka, 2009, create by self

In figure 12, there is a linear positive correlation between flip angle and SNR. Like previous studies, this is a result of the increase in magnitude of transverse magnetic field after excitation.

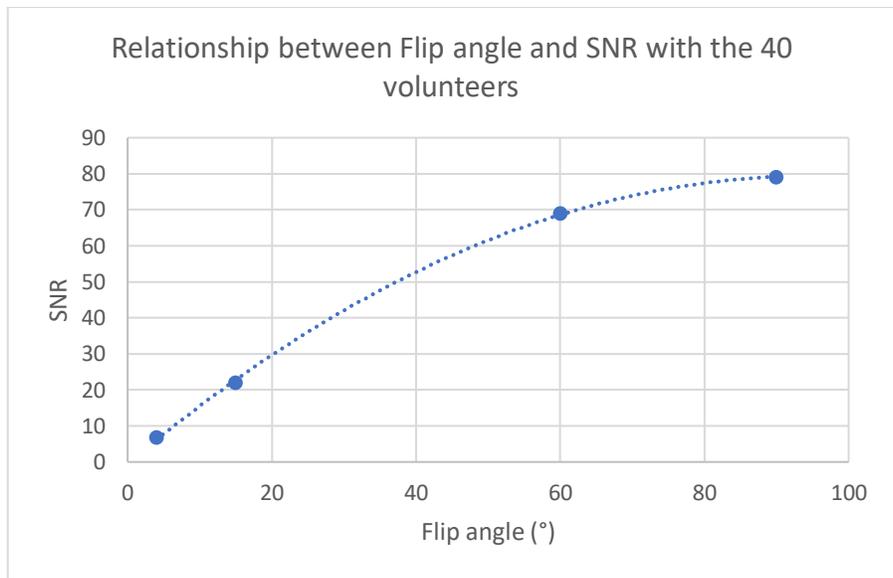

Figure 13. Dataset by Tanaka, 2009, create by self

Figure 13 illustrates a non-linear positive correlation between flip angle and SNR. As the flip angle increases, the SNR also increases. The range of SNR varies from 6.7 to 79 (Tanaka, 2009), representing a substantial percentage increase of 1080%. These findings highlight the significant impact of the flip angle on SNR.

The non-linear tendency observed in this study aligns with the findings of study 1, which is expected since blood, regardless of the imaging location, is imaged. This suggests that the T1 value of the blood would be comparable in both studies. Additionally, the limit of proportionality in both studies is

Acronyms:

*M:* net magnetization, $M_{xy}$: strength and direction of transverse magnetic field, $M_z$: strength and direction of longitudinal magnetic field, $B_0$: external background magnetic field strength, TR: repetition time, SNR: signal-to-noise ratio remarkably similar, approximately 40° in study 1 and around 35° in this study. This similarity supports the assertion of similar nuclear properties of blood across different body locations.

Despite the relatively high T1 value of 1350ms for oxygenated blood at a magnetic field strength of 1.5T (as observed in figure 10), it is unexpected that, in practice, a larger flip angle is found to be optimal, in line with study 1. This contradicts the initial expectation of a small optimum flip angle suggested by the Ernst angle equation.

*Study 4:*

The objective of this study is to optimize the imaging protocol for Susceptibility-Weighted Imaging (SWI), which utilises a 3D Gradient Echo (GRE) sequence to capture images of the brain and small blood vessels within it. To achieve this, a correlation has been established between a raw magnitude image of the brain and a simulated curve representing the signal intensity against flip angle for different types of tissue:

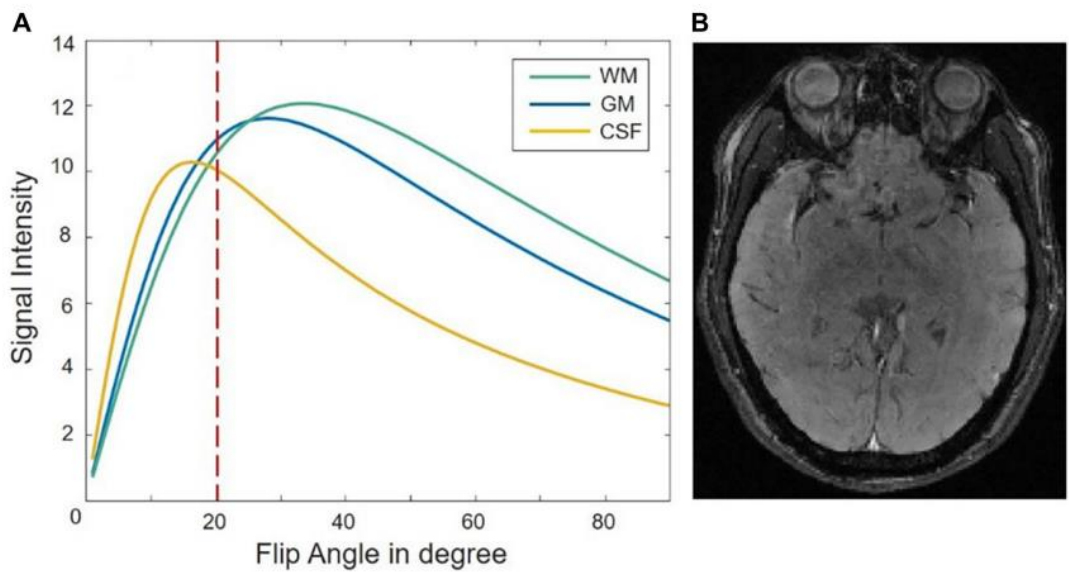

Figure 14, (A) a simulation curve of signal intensity against flip angle, where WM stands for white matter, GM stands for grey matter, CSF stands for Cerebrospinal fluid (B) a brain MRI image at a flip angle of 20°, by Qiu, 2015

Signal intensity in MRI images determines the brightness of structures, with higher intensity indicating brighter regions and lower intensity corresponding to darker areas. Figure 14A confirms that grey matter exhibits the highest signal intensity, followed by white matter and cerebrospinal fluid. The distinction between grey and white matter intensities is relatively small, consistent with the findings of figure 14B, which shows dark regions between the skull and brain, representing cerebrospinal fluid. Additionally, the slightly darker appearance of the brain's centre, as observed, supports figure 15, illustrating the presence of white matter in the inner regions. Thus, the simulation curve employed in this study accurately represents these signal intensity patterns.

Acronyms:

*M*: net magnetization, $M_{xy}$: strength and direction of transverse magnetic field, $M_z$: strength and direction of longitudinal magnetic field, $B_0$: external background magnetic field strength, TR: repetition time, SNR: signal-to-noise ratio

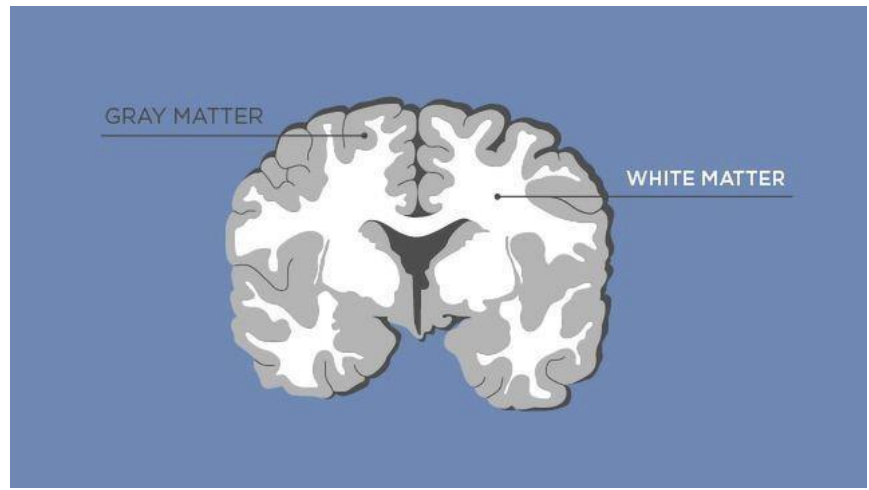

Figure 15, a diagram of brain, MacKenzie, 2019

Figure 14A reveals that the curves representing WM, GM, and CSF exhibit similar shapes but differ in their maximum signal intensities. Notably, the T1 values depicted in figure 16 align with this observation, as WM possesses the lowest T1 (493ms), followed by GM (717ms), and CSF with the highest T1 (2,200ms). These findings support the trends observed in figure 14A, which demonstrate the variation of Ernst angles for these tissue types. Specifically, the graph indicates Ernst angles of approximately 17°, 36°, and 23° for CSF, WM, and GM, respectively, aligning with the theoretical Ernst angles based on their respective T1 values. Consequently, the curve in this study corroborates the findings of figure 9.

| Magnetic field (T) | Tissue | T1 (ms) | T2* (ms) |
|---|---|---|---|
| 0.5 | White matter | 493 | 72 |
| | Gray matter | 717 | 86 |
| | Cerebrospinal fluid | 2,200 | 200 |
| 1.5 | White matter | 780 | 64 |
| | Gray matter | 920 | 83 |
| | Cerebrospinal fluid | 2,500 | 100 |

Figure 16, a table with t1 values of different tissues under 0.5T and 1.5T

### Evaluation

*Significance of SNR enhancement independent of Flip angle*

Studies 1, 2, and 3 have demonstrated varying levels of significance in SNR improvement. Study 3 exhibited the highest percentage increase of 1080%, followed by study 1 with a 500% increase, and finally study 2 with a 20% increase. Despite these variations, the ranges of flip angles used differed between the studies. To facilitate comparison between SNR and other parameters independent of flip angle, it is helpful to consider the average percentage increase per degree for each study. These values are approximately 6% for study 1, 1% for study 2, and 13% for study 3. This calculation is valid as the trends between flip angle and SNR remain consistent.

| Study 1 | Study 2 | Study 3 | Study 4 |
|---|---|---|---|
| Imaging type: Phase Contrast (PC) MRI | Imaging type: Dynamic contrast enhanced (DCE) MRI | Imaging type: Phase Contrast (PC) MRI | Imaging type: Susceptibility weighted imaging (SWI) |
| Body part: leg vessels (blood) | Body part: liver tissue | Body part: neck vessels (blood) | Body part: brain |
| Magnetic field: 1.5T | Magnetic field: 1.5T | Magnetic field: 1.5T | Magnetic field: 1.5T |
| Imaging parameters: Field of view: 340 x 212.5 mm Matrix: 256 x 160 Slice thickness: 8mm | Imaging parameters: Field of view: 271 x 379 mm Matrix: 448 x 224 Slice thickness: 5mm | Imaging parameters: Field of view: 200 mm Matrix: 256 x 256 Slice thickness: 0.8mm | Imaging parameters: N/A for relevance: simulated curve |

Figure 17, a table with information of studies, by self, from (Ennis, 2011), (Lee, 2015), (Tanaka, 2009), (Qiu, 2022)

According to figure 17, studies 1 and 3 both focus on imaging blood, whereas study 2 targets liver tissue. Therefore, the percentage increase per degree in studies 1 and 3 may generally be higher than in study 2, potentially due to differences in the imaged sample type. Additionally, the field of view is greatest in study 3, followed by study 1, and finally study 2. This suggests that the percentage increase per degree could be influenced by the size of the field of view, with a smaller field of view potentially resulting in a higher rate of signal intensity increase and hence SNR improvement. In addition, no clear correlation is evident between matrix and slice thickness, and the rate of percentage increase. However, it is important to note that the rate of percentage increase may not be solely determined by the imaging parameters discussed, as there are likely other factors at play that were not addressed in the paper.

Altogether, the range of percentage increase per degree across the three studies is 12%, which is lower than the smallest percentage increase of 20% in study 2 resulting from the flip angle. This emphasizes the significant impact of flip angle on SNR and image quality, irrespective of variations in the rate of percentage increase due to other imaging parameters.

*Consistency with Theoretical Expectations*

The observed patterns in SNR or signal intensity with respect to flip angle (shown in figures 8, 11, 13, and 14A) exhibit similarities. However, studies 1, 2, and 3 present unexpected optimal flip angles when considering the Ernst angle as the optimal angle. This discrepancy can be attributed to the pulse sequence used in these studies. According to Elster (2023), the Ernst angle maximises signal intensity specifically when a spoiled GRE sequence is used. This sequence destroys the remaining transverse magnetic field after digitizing the MRI signal (Bushong, 2015, pp.282). Reasonably, none of the studies utilized a spoiled GRE sequence, except for study 4. Additionally, the derivation of the Ernst angle assumes a steady-state condition during imaging. Consequently, it can be concluded that studies 1, 2, and 3 do not satisfy the steady-state condition. Therefore, only study 4, which optimized signal intensity using the Ernst angle, aligns with expectations as it used a spoiled GRE sequence in steady state condition. Therefore, Ernst angle optimization has proven to be an inconsistent optimization approach, thus it is

Acronyms:

*M*: net magnetization, M$_{xy}$: strength and direction of transverse magnetic field, M$_z$: strength and direction of longitudinal magnetic field, B$_0$: external background magnetic field strength, TR: repetition time, SNR: signal-to-noise ratio crucial to conduct experiments under varying sets of imaging parameters to achieve flip angle optimization.

## Discussion

Flip angle plays a vital role in maximising signal intensity and SNR in an image. However, it is important to consider other factors that can also impact SNR, such as the background magnetic field. For example, increasing the background magnetic field from 3T to 7T can result in a 63% increase in SNR (Zhang, 2013). Additionally, equipment-related factors like voxel size, RF coil sensitivity, RF coil size and selection, and receiver bandwidth should be considered when aiming to further enhance SNR, even after the flip angle has been optimized.

A high SNR in an image does not guarantee its accuracy or quality for clinical diagnosis, as highlighted by Tanaka (2009). While SNR increases with higher flip angles (until reaching optimum angle), it comes with the drawback of increased uncertainties and a greater potential for overestimation due to the partial volume effect (Tanaka, 2009). This effect causes blurring of moving structures as they span multiple image voxels during data acquisition. Consequently, a higher SNR does not necessarily translate to a better image. It is crucial to strike a balance and adjust the flip angle within a certain range that ensures both an acceptable SNR and image accuracy.

Even if the system is in a steady state, as mentioned by Qiu (2022), and the signal intensity can be optimized using the Ernst angle, it is important to consider that different tissues have different Ernst angles. Therefore, selecting the Ernst angle of one tissue alone is not sufficient. The chosen angle must provide adequate contrast for the specific targeted tissue.

As previously mentioned, image quality in MRI relies on various factors, including the CNR, which compares the tissue of interest to the background noise. A higher CNR indicates better distinguishability of the tissue. This can be observed in the central diagram of figure 18:

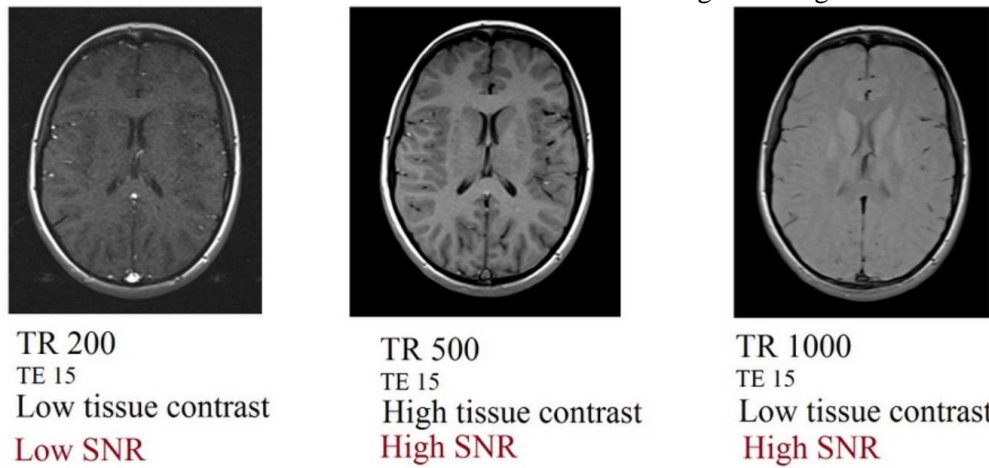

TR 200
TE 15
Low tissue contrast
Low SNR

TR 500
TE 15
High tissue contrast
High SNR

TR 1000
TE 15
Low tissue contrast
High SNR

Figure 18, a comparison between images of high and low CNR and SNR, by mrimasters, 2023.

Achieving high image resolution is also essential to ensure clear visualization of structures. Therefore, it is crucial to conduct additional investigations to explore the correlation between flip angle and other indicators of image quality, such as Contrast-to-Noise Ratio (CNR) and resolution as well as between other imaging parameters (echo time, TR) and equipment related factors (voxel size, receiver bandwidth, $B_0$, etc) and indicators of image quality

Acronyms:

$M$: net magnetization, $M_{xy}$: strength and direction of transverse magnetic field, $M_z$: strength and direction of longitudinal magnetic field, $B_0$: external background magnetic field strength, TR: repetition time, SNR: signal-to-noise ratio

## Conclusion

In conclusion, it can be inferred that the flip angle serves as an essential parameter in determining the SNR and image quality, owing to the fact that the increase in SNR with respect to the change in flip angle per degree is relatively low when juxtaposed with the overall percentage increase in SNR. The available studies indicate a non-linear correlation between flip angle and SNR. Specifically, an increase in flip angle is associated with a corresponding increase in SNR up to a certain point where an optimal angle is achieved. This is due to a higher amount of $M$ being placed into the XY plane, resulting in an increase in $M_{xy}$ and subsequent signal generation. However, once the optimal value is reached, the SNR experiences a decline as the flip angle continues to rise. This can be attributed to the diminishing suitability of $M_z$ for subsequent stimulations. As such, achieving the ideal flip angle necessitates striking a balance between both rationales.

The implementation of Ernst angle optimization methodology is applicable solely to the utilization of spoil GRE sequence in a state of steady equilibrium. Given the inconsistent nature of Ernst angle optimization, it becomes essential to conduct experiments with diverse imaging parameters in order to achieve optimal flip angle optimization.

Acronyms:

$M$: net magnetization, $M_{xy}$: strength and direction of transverse magnetic field, $M_z$: strength and direction of longitudinal magnetic field, $B_0$: external background magnetic field strength, TR: repetition time, SNR: signal-to-noise ratio

## Appendix

*Derivation of Ernst Angle*

According to Brown (2014),

$$\frac{d\vec{M}}{dt} = \gamma \vec{M} \times \vec{B}$$

Where $\vec{M}$ refers to the direction of net magnetisation, $\gamma$ refers to the gyromagnetic ratio, and $\vec{B}$ refers to the direction of magnetic field, that aligns with the Z axis. Therefore, $B = B_0\,\hat{z}$. Then, as indicated by Colman (2020), "In the absence of the interactions between spins (no relaxation) the three equations generated by the cross product are:" (pp.2)

$$\frac{dM_x}{dt} = \gamma M_y B_0 \hat{x} = \omega_0 M_y$$

$$\frac{dM_y}{dt} = -\gamma M_x B_0 \hat{y} = -\omega_0 M_x$$

$$\frac{dM_z}{dt} = 0$$

Acronyms:

*M*: net magnetization, M$_{xy}$: strength and direction of transverse magnetic field, M$_z$: strength and direction of longitudinal magnetic field, B$_0$: external background magnetic field strength, TR: repetition time, SNR: signal-to-noise ratio

Where $M_x$, $M_y$, and $M_z$ refers to magnetisation of the X, Y, and Z axis respectively, and $\omega_0$ refers to the nuclei's precessional frequency. Using Wolfram Mathematica (Wolfram Research, Inc., 2022) to solve the Bloch equations, a set of solution is derived (check code snippets):

$$M_x(t) = M_x(0)cos(\omega_0)t + M_y sin(\omega_0)t$$

$$M_y(t) = M_y(0)cos(\omega_0)t - M_x sin(\omega_0)t$$

Then, to add in relaxation, starting with $M_z$, after excitation, the recovery rate of this component will be proportional to the difference between equilibrium, $M_0$, and $M_z$ (Colman, 2020, pp.2):

$$\frac{dM_z}{dt} = \frac{1}{T_1}(M_0 - M_z)$$

Where $T_1$ refers to the time of $M_z$ to recover to steady state, $T_1$ relaxation time. Similarly, a function subject to $M_z$ is derived by Wolfram Mathematica:

$$M_z(t) = M_z(0)e^{-t/T_1} + M_0(1 - e^{-t/T_1})$$

And verified by......

And similarly, adding the relaxation of $M_x$ and $M_y$ at rate $\frac{1}{T_2}$ (Olman, 2020):

$$\frac{dM_x}{dt} = \omega_0 M_y - \frac{1}{T_2}M_x$$

$$\frac{dM_y}{dt} = \omega_0 M_x - \frac{1}{T_2}M_y$$

Where $T_2$ refers to the time of $M_{xy}$ to recover to steady state, $T_2$ relaxation time. Similarly, Wolfram Mathematica provided derivation of two functions subject to $M_x$ and $M_y$:

$$M_x(t) = e^{-t/T_2}(M_x(0)cos(\omega_0)t + M_y(0)sin(\omega_0)t)$$

$$M_y(t) = e^{-t/T_2}(M_y(0)cos(\omega_0)t - M_x(0)sin(\omega_0)t)$$

Which can be further combined to:

$$M_{xy} = M_{xy}(0)e^{-t/T_2}$$

In a simple inversion recovery sequence, a pulse sequence uses a 90° pulse following a 180° inversion pulse (Bushong, 2015, pp.62), which flips $M$ to $-M_0$. It should be noticed that regardless $M_{xy}$ induces the signal, it is the magnitude of $M_z$ that decides the signal intensity, as the size of $M_{xy}$ is directly dependent of $M_z$ as $M_z$ is tipped into the transverse plane after excitation. If TR is a lot greater than $T_1$ and $T_2$, then $M_z$ straight after the inversion pulse:

$$M_z(TI^+) = M_z(TI^-)\cos(\alpha)$$

Where $\alpha$ is the RF pulse's flip angle.

Then, at t = TE, the time after excitation for data acquisition, $M_z$:

$$M_z(TE) = M_z(TI^+)e^{-TE/T_1} + M_0(1 - e^{-TE/T_1})$$

In general, $M_z$ after reaching steady state, where $M$ is at equilibrium as it receives a series of RF pulses (Miller, 2011, pp.93), at the n[th] pulse sequence ( t $= n \times TR$):

$$M_z(t_{n+1}) = M_z(t_n)cos\ (\alpha)e^{-TR/T_1} + M_0(1 - e^{-TR/T_1})$$

It should be noticed that since $M_z$ have reached steady state, $M_z(t_{n+1})$ would therefore equal to $M_z(t_n)$, which is equivalent to M$_{ss}$. By substitution,

$$M_z(t_{n+1}) = M_z(t_n)cos(\alpha)e^{-\frac{TR}{T_1}} + M_0(1 - e^{-\frac{TR}{T_1}})$$

$$M_{ss} = M_{ss}cos\ (\alpha)e^{-TR/T_1} + M_0(1 - e^{-TR/T_1})$$

$$M_{ss}(1 - cos\ (\alpha)e^{-TR/T_1}) = M_0(1 - e^{-TR/T_1})$$

And hence:

$$M_{ss} = \frac{M_0(1 - e^{-TR/T_1})}{1 - cos(\alpha)e^{-TR/T_1}} \qquad \boxed{1}$$

Resolving $M_{ss}$ to the transverse plane:

$$M_z\ (TR^+) = M_{ss}sin(\alpha)$$

Substituting $\boxed{1}$ :

$$M_z\ (TR^+) = \frac{M_0(1 - e^{-TR/T_1})}{1 - cos(\alpha)e^{-TR/T_1}} sin(\alpha)$$

Which is the relationship between flip angle and signal intensity. The resultant solution of $\alpha$ when derivate is zero is derived as:

$$\alpha = cos^{-1}(e^{-TR/T_1})$$

Also formally known as the Ernst Angle.

Acronyms:

$M$: net magnetization, M$_{xy}$: strength and direction of transverse magnetic field, M$_z$: strength and direction of longitudinal magnetic field, B$_0$: external background magnetic field strength, TR: repetition time, SNR: signal-to-noise ratio

**Code Snippets**

*Solving Bloch Equation*

```
eqns = {
    mx'[t] == ω0 * my[t],
    my'[t] == -ω0 * mx[t],
    mz'[t] == 0
};
sol = DSolve[eqns, {mx[t], my[t], mz[t]}, t];
sol
```

Out[●]= $\{\{mx[t] \to c_1 \cos[t\,\omega0] + c_2 \sin[t\,\omega0], my[t] \to c_2 \cos[t\,\omega0] - c_1 \sin[t\,\omega0], mz[t] \to c_3\}\}$

*Solving for longitudinal relaxation (T1)*

```
eqns = {mz'[t] == (1 / T1) * (m0 - mz)};
sol = DSolve[eqns, mz, t];
sol
```

Out[●]= $\left\{\left\{mz \to \text{Function}\left[\{t\}, m0 + e^{-\frac{t}{T1}} c_1\right]\right\}\right\}$

*Verify as t -> ∞*

```
sol = mz[t] /. DSolve[{mz'[t] == (1 / T1) * (m0 - mz[t])}, mz, t][[1]]
limitSol = Limit[sol, t → Infinity]
```

Out[●]= $m0 + e^{-\frac{t}{T1}} c_1$

Out[●]= $m0 \quad \text{if} \quad (m0 \mid c_1) \in \mathbb{R} \,\&\&\, T1 > 0$

*Solving for transverse relaxation (T2)*

```
eqns = {
    mx'[t] == ω0 * my[t] - (1 / t2) * mx[t],
    my'[t] == -ω0 * mx[t] - (1 / t2) * my[t]
};
sol = DSolve[eqns, {mx[t], my[t]}, t];
sol
```

Out[●]= $\left\{\left\{mx[t] \to e^{-\frac{t}{t2}} c_1 \cos[t\,\omega0] + e^{-\frac{t}{t2}} c_2 \sin[t\,\omega0], my[t] \to e^{-\frac{t}{t2}} c_2 \cos[t\,\omega0] - e^{-\frac{t}{t2}} c_1 \sin[t\,\omega0]\right\}\right\}$

Acronyms:

*M*: net magnetization, $M_{xy}$: strength and direction of transverse magnetic field, $M_z$: strength and direction of longitudinal magnetic field, $B_0$: external background magnetic field strength, TR: repetition time, SNR: signal-to-noise ratio

Acronyms:

*M:* net magnetization, $M_{xy}$: strength and direction of transverse magnetic field, $M_z$: strength and direction of longitudinal magnetic field, $B_0$: external background magnetic field strength, TR: repetition time, SNR: signal-to-noise ratio